# Structure of the Universe: Generalization of Weinberg relation

**A Alfonso-Faus**
E.U.I.T. Aeronáutica, Plaza Cardenal Cisneros s/n, 28040 Madrid, SPAIN
E-mail: aalfonsofaus@yahoo.es

**Abstract.** Weinberg relation can be explained by Newton laws. It can also be transformed into a relation between a gravitational cross section area, associated to a gravitational mass m, and the product of two gravitational radii corresponding to the mass m and to the mass M of the Universe. Using a physical similarity process we propose a convective cascade mechanism to explain the structures present in the Universe. The proposal includes the vacuum fluctuations as the heating source for convection to take place. The fundamental Rayleigh number is used with a value of 1100, as in the case of the granules observed in the sun photosphere. Computer simulations results give structures that resemble a configuration of the type of the Benard cells. We present the similarity of our results with the actual structures in the Universe (masses and sizes).

## 1. – INTRODUCTION

It is well known that simple combinations of physical constants may give significant new insights in theoretical physics. For example combining Planck's constant ħ (quantum mechanics), the speed of light c (relativity) and the gravitational constant G (gravitation), gives a type of units of mass, length and time that can be related to vacuum fluctuations. If we include a cosmological parameter, like the Hubble H, Weinberg (1972) found an order of magnitude for the mass of fundamental particles. We show here that this relation can be derived from Newton's laws. It also can be transformed into a law relating a gravitational cross section area, a function of the mass m, and its gravitational radius.

We then generalize Weinberg's relation to any mass that has universal existence. This means that their gravitational interaction is complete, i.e., its action is confined to the size of the Universe, no more no less. And this is as much as to say that the "range" of the gravitational interaction is of the order of magnitude of the size of the Universe ct. We go on proposing a physical similarity between the convection cells, a Bénard configuration, and the structures in the Universe. We consider a gas in the Universe that is heated by the vacuum fluctuations. The condition for having convection is given by the value of the Rayleigh number Ra > 1100. A convective cascade is postulated and the resultant structures found. The similarity of this result with the actual structures in the Universe is presented and found to be in striking agreement.

## 2. – WEINBERG'S RELATION

In 1972 Weinberg discussed a relation derived by using G (gravitation), c (relativity), ħ (quantum mechanics) and the Hubble parameter H (cosmology) to arrive at the order of magnitude of the mass m of a fundamental particle:

$$m^3 \approx \frac{\hbar^2 H}{Gc} \qquad (1)$$

Following a Machean approach, Alfonso-Faus (2008), the maximum momentum content of a mass m' was considered to be m'c, a constant due to the rest of the Universe. If this momentum has been acquired during a cosmological time t, then the average force that can exert is f = m'c/t. This average force f can be thought of as an action fluxing over a spherical surface of radius r centred at the centre of gravity of m'. Consider it to act upon a small test particle m which has an effective interaction area $\sigma_m$. Then m will feel a force exerted by the presence of m' given by

$$f = \frac{m'c}{t}\frac{\sigma_m}{4\pi r^2} \qquad (2)$$

There are reasons to believe that the gravitational field may be treated as a tachyon field: The action of gravity may well travel at a speed much higher than the speed of light c, Van Flandern (1998), Alfonso-Faus (2002). Since tachyons travel backwards in time, the force in (2) is negative. By equating this force to the gravitational attractive force between m' and m we get:

$$-\frac{Gm'm}{r^2} = f = -\frac{m'c}{t}\frac{\sigma_m}{4\pi r^2} \qquad (3)$$

and rearranging we arrive at

$$\sigma_m = 4\pi\frac{Gm}{c^2}. \quad ct \qquad (4)$$

We can say that the gravitational interaction area $\sigma_m$ of a mass m is of the order of the product of its gravitational radius, $Gm/c^2$, times the size of the Universe ct. We will later present a strong evidence for the area $\sigma_m$ to be of the order of the "geometrical" (gravitational) cross section of the mass m. Following this idea we can take the square of the Compton wavelength, given by ħ/mc for a fundamental particle, as the approximate value of $\sigma_m$. Substituting this expression in (4) we get:

$$4\pi m^3 = \frac{\hbar^2}{Gct} \qquad (5)$$

and taking the value for the Hubble parameter $H \approx 1/t$ we finally get the relation (1) advanced by Weinberg. We have proved, Alfonso-Faus (2008), that this is the case and that ct is constant. Relation (1) has the important property that it consists of a combination of quantum and cosmological parameters. And it is the result of Newton's laws. The order of magnitude of the mass of the fundamental particles is then a result of Newton's laws. We have used a Machean approach when considering the mass m', and its momentum m'c, as due to the presence of the rest of the Universe. Now we see that we have obtained one more thing: the quantity of mass in a fundamental particle is explained by Newton's laws. Since cosmology (H), gravity (G) and quantum mechanics (ħ), are built in Weinberg's relation (1) that can be explained by these laws, this strongly suggests that they may have a deeper meaning than usually thought in order to understand Nature, as we will see.

## 3. - FORMATION OF STRUCTURES FROM CLUSTERS OF GALAXIES TO FUNDAMENTAL PARTICLES

The gravitational cross section area found in (4) can explain the formation of all structures in the Universe, from the clusters of galaxies to the fundamental particles, as a gravitational effect. It is a geometric inversion law that gives the gravitational sizes of all structures in the Universe. If one introduces the actual size and mass of structures like the whole Universe, clusters of galaxies, galaxies, globular clusters, nebulae, and even the fundamental particles, one can check that the relation (4) is very well satisfied by all these structures (see Table 1). The conclusion is that Newton's laws are at the base of the explanation of the formation of all different structures in the Universe. This is something that has been always considered to be the case, but now we have more evidence for it.

## 4. – GENERALIZATION OF WEINBERG'S RELATION

The geometric relation (4) that we have found is in fact a generalization of Weinberg's relation (1). When using Compton's wavelength we get this particular result. However, we can generalize it by considering that the gravitational cross section area $\sigma_m$ may have any value. To keep the Machean approach that we have followed we consider only masses m that have a universal existence. This means that the number N of masses m existing in the Universe has to give a total mass Nm of the order of the total mass M of the Universe itself (about $10^{56}$ grams). Then, multiplying (4) by N we get:

$$N\sigma_m = 4\pi \frac{GNm}{c^2}.\ ct = 4\pi \frac{GM}{c^2}.\ ct \approx (ct)^2 \tag{6}$$

Considering a Universe of size ct then its cross sectional area is of the order of $(ct)^2$ and from (6) we see that this area is completely covered by the N objects each having a gravitational cross section $\sigma_m$ projected into a plane. It is a geometric interpretation in two dimensions of the principle that we call "complete universal interaction", in this case gravitational. The interaction covers the whole Universe. It has a range just to fill it, no more no less. In a very simple classical way one can say that the mean free path of this interaction is just the size of the universe:

$$ct \approx \frac{1}{n_m \sigma_m} = \frac{(ct)^3}{N} \cdot \frac{1}{\sigma_m} \tag{7}$$

this is the same as (6). This two dimensional picture is obtained by projecting all the gravitational cross sections $\sigma_m$ of all the masses m in the Universe on a plane. The resultant area is $(ct)^2$, which is the gravitational cross section of the Universe.

## 5. – STRUCTURE OF THE UNIVERSE: THE CONVECTIVE CASCADE

The sun's photosphere is full of granules of average size $10^8$ cm in diameter. This means that the photosphere has about $2x10^6$ granules in its surface. The dimensionless number of Rayleigh, $R_a$ , that defines the starting of convection to get the Bénard configuration, has the critical value of 1100 for the case of a free top surface heated from below. Natural convection in fluid layers heated from below is a classical subject, Chandrasekhar (1961), Turner (1973). Since the Rayleigh number is of the order of magnitude of the ratio of the size of the container to the size of each cell, the ratio of areas is of the order of $Ra^2$. We then see that, for the case of the sun's photosphere, the condition for convection is fully satisfied. The number $10^6$ is the order of magnitude for the number of cells in a Bénard configuration. Computer simulations results present structures that resemble a configuration of the Bénard cells type, Narlikar (2002).

We now consider that, at the initial stages of the Universe, we have a uniform gas filling it. We also have vacuum fluctuations. Now we postulate a physical similarity between the two dimensional plate heating a gas on it, with the condition for convection fulfilled, and the two dimensional projection of the Universe on a plane giving an area $(ct)^2$. The vacuum fluctuations are the source of heat. Convection takes place, giving granules of size $\sigma_m$, when $R_a^2 \approx 10^6 = N/ \sigma_m$. As soon as each object $\sigma_m$ is formed it can be considered as an individual one undertaking another convective phenomena. And so on. After a number n of such convective events, forming a cascade, we have a number of $N = 10^{6n}$ objects in the whole Universe, each one with a gravitational size $\sigma_m$ The postulated physical similarity implies a *geometrical and process-related similarity* between the gas on the heated plate and the gas heated by the vacuum fluctuations in the Universe. Then we have a law of structures in the Universe given by the number n = 0, 1, 2, 3, 4, 5, 6, 7, 8, 9, 10, 11, 12, 13, 14, 15, 16, 17, 18, 19, and 20, that defines the number, mass and size of each object in a hierarchical way. Table 2 gives for these values of n the number of objects of each class, their mass and their linear size. It is seen that the law of structures, from n = 0 to n = 5, needs the mass to be multiplied by the factor 100 in order to have a complete match from n = 0 to n = 20. We interpret this as a result of the presence of dark matter and/or dark energy in the first 6 structures.

## 6. – CONCLUSIONS

Weinberg's relation can be explained by Newton's laws. It can be transformed to introduce the concept of a gravitational cross section area. Postulating a convective-cascade mechanism, it can be generalized from fundamental-particle mass prediction to all the universal structures. It gives a hierarchical law of structures in the Universe. Out of the 21 predicted hierarchical structures we have identified 13. The remaining 8 have to be analyzed.

## 7. – REFERENCES

**TABLE 1 Astrophysical Objects**

| Object | Mass | Gravitational radius $r_g$ | $r_g$ .ct | Size $r = (r_g ct)^{1/2}$ | Actual size |
|---|---|---|---|---|---|
| Universe | $10^{56}$ grams | ct = $10^{28}$ cms | $10^{56}$cm$^2$ | $10^{28}$ cms | $10^{28}$ cms |
| Galactic cluster | $10^{49}$ grams | $10^{21}$ cms | $10^{49}$cm$^2$ | $3x10^{24}$ cms | $10^{24}$ cms |
| Galaxies | $10^{44}$ grams | $10^{16}$ cms | $10^{44}$cm$^2$ | $10^{22}$ cms | $10^{22}$ cms |
| Globular clusters | $10^{38}$ grams | $10^{10}$ cms | $10^{38}$cm$^2$ | $10^{19}$ cms | $10^{19}$ cms |
| Nebulae | $2x10^{38}$grams | $2x10^{10}$ cms | $2x10^{38}$cm$^2$ | $1.4x10^{19}$ cms | $10^{19}$ cms |
| Protostar | $2x10^{33}$grams | $2x10^{5}$ cms | $2x10^{33}$cm$^2$ | $4.5x10^{16}$ cms | $10^{16}$ cms |
| Protons | $1.7x10^{-24}$grams | $1.7x10^{-52}$ cms | $1.7x10^{-24}$cm$^2$ | $1.3x10^{-12}$cms | $10^{-12}$ cms |

From $r = 10^{-12}$ cms (protons) up to $r = 10^{28}$ cms (Universe) the geometric inversion implied by the relation $\sigma_m = 4\pi\ Gm/c^2$ . ct is remarkably well satisfied by the Universe, cluster of galaxies, galaxies, globular clusters, nebulae, protostar systems and protons.

**TABLE 2 Structure of the Universe**

| n | $N = 10^{6n}$ | Mass M/N | Size $(\sigma_m)^{1/2}$ | Object |
|---|---|---|---|---|
| 0 | 1 | (x100) $10^{56}$ grams | $10^{28}$ cms | Universe |
| 1 | $10^{6}$ | (x100) $10^{50}$ grams | $10^{25}$ cms | Cluster of galaxies |
| 2 | $10^{12}$ | (x100) $10^{44}$ grams | $10^{22}$ cms | Galaxies |
| 3 | $10^{18}$ | (x100) $10^{38}$grams | $10^{19}$ cms | Globular clusters |
| 4 | $10^{24}$ | (x100) $10^{32}$grams | $10^{16}$ cms | Nebulae |
| 5 | $10^{30}$ | (x100) $10^{26}$ grams | $10^{13}$ cms | Planet (diluted) |
| 6 | $10^{36}$ | $10^{18}$ grams | $10^{9}$ cms | Comets |
| 7 | $10^{42}$ | $10^{12}$ grams | $10^{6}$ cms | |
| 8 | $10^{48}$ | $10^{6}$ grams | $10^{3}$ cms | Meteorites |
| 9 | $10^{54}$ | 1 gram | 1 cm | Water drop |
| 10 | $10^{60}$ | $10^{-6}$ grams | $10^{-3}$ cms | Whisker |
| 11 | $10^{66}$ | $10^{-12}$ grams | $10^{-6}$ cms | |
| 12 | $10^{72}$ | $10^{-18}$ grams | $10^{-9}$ cms | |
| 13 | $10^{78}$ | $10^{-24}$ grams | $10^{-12}$ cms | Protons |
| 14 | $10^{84}$ | $10^{-30}$ grams | $10^{-15}$ cms | |
| 15 | $10^{90}$ | $10^{-36}$ grams | $10^{-18}$ cms | Photon number and mass |
| 16 | $10^{96}$ | $10^{-42}$ grams | $10^{-21}$ cms | |
| 17 | $10^{102}$ | $10^{-48}$ grams | $10^{-24}$ cms | |
| 18 | $10^{108}$ | $10^{-54}$ grams | $10^{-27}$ cms | |
| 19 | $10^{114}$ | $10^{-60}$ grams | $10^{-30}$ cms | |
| 20 | $10^{120}$ | $10^{-66}$ grams | $10^{-33}$ cms | Gravity quanta mass and Planck's size |

From $r = 10^{-33}$ cms (Planck's size) up to $r = 10^{28}$ cms (Universe) the geometric inversion given by $\sigma_m = 4\pi\ Gm/c^2.ct$ and the convective cascade law ($N = 10^{6n}$) are remarkably well satisfied by the structures in the Universe.